# Topology Learning of Radial Dynamical Systems with Latent Nodes

Saurav Talukdar[1], Deepjyoti Deka[2], Michael Chertkov[2] and Murti Salapaka[3]

*Abstract*— In this article, we present a method to reconstruct the topology of a partially observed radial network of linear dynamical systems with bi-directional interactions. Our approach exploits the structure of the inverse power spectral density matrix and recovers edges involving nodes up to four hops away in the underlying topology. We then present an algorithm with provable guarantees, which eliminates the spurious links obtained and also identifies the location of the unobserved nodes in the inferred topology. The algorithm recovers the exact topology of the network by using only time-series of the states at the observed nodes. The effectiveness of the method developed is demonstrated by applying it on a typical distribution system of the electric grid.

## I. INTRODUCTION

Networks provide a convenient framework for analysis of the behavior of large scale systems and have found applications in neuroscience [1], biology [2], finance [3] and many more. In this regard, an important question of interest is to estimate the unknown topology or the interaction structure amongst the entities of the network using measurements. There are both active [4] and passive approaches [5] to infer the unknown influence structure/ topology from measurements. Active approaches include performing interventions on the network, which are not always possible. For example: it is not viable to perform experiments by removing links in a power distribution network. In this article we are interested in passive approaches to topology learning from observed time series measurements.

The problem of topology or structure learning under the assumption that the entities are random variables is an active area of research for the past few decades and a good summary is available in [6], [7], [8], [9]. However, the random variables framework does not capture the dynamics amongst the entities and hence is not useful for application where lagged dependencies are present. Such applications include power distribution networks, seasonality in climate systems [10], finance, thermal dynamics of buildings [11] and many more.

Recently, there has been considerable interest in the topology learning for linear dynamical systems from time series measurements. It this regard some of the notable works are [12], [13], [14], [15], [16], [17]. However, these works assume that all nodes in the network are observed or full

[1] Saurav Talukdar is with Department of Mechanical Engineering, University of Minnesota, Minneapolis, USA, sauravtalukdar@umn.edu
[2] Deepjyoti Deka and Michael Chertkov are with Los Alamos National Lab, Los Alamos, USA, deepjyoti,chertkov@lanl.gov
[3] Murti V. Salapaka is with Department of Electrical and Computer Engineering, Minneapolis, USA, murtis@umn.edu

The authors acknowledge the support of Department of Energy's ARPA-E program, the Grid Modernization Lab Consortium, and the Center for Non Linear Studies (CNLS) at Los Alamos for this work.

network observability. Topology reconstruction from passive measurements for a network of linear dynamical systems with unobserved nodes is discussed in [18], [19]. The problem formulation in [18] is focused on directed poly-tree network of linear dynamical systems with unobserved nodes. The framework presented in [19] is restricted to Gaussian stationary time series and does not include consistency of identification results. In this article, we present an algorithm which leads to recovering the exact topology of the network under partial observation of the nodes. In particular, we are interested in radial linear dynamical systems, which are characterized by a tree topology with undirected (that is bi-directed) edges between neighbors rather than uni-directed edges. Indeed the directed loops are a part of the framework presented here unlike [18].

Radial linear dynamical systems (RLDS) [20] represent an important class of networks in engineering systems. Among others, RLDS can model dynamics in power distribution systems. An algorithm for exact topology learning for RLDS with all nodes being observed is presented in [20]. We show that for RLDS, under the assumption that unobserved nodes are 'deep into the network' such that their effects are felt through observed nodes, it is possible to recover the underlying interaction topology exactly. In this regard, we build upon topological separation ideas of [20] and phase response properties of [21], to devise an algorithm which provably recovers the exact topology of the RLDS. Our algorithm uses only the time series measurements from the nodes and does not use any knowledge of system parameters as well as the exogenous injections. The efficacy of the algorithm is demonstrated on a 39 bus radial power network with linearized swing dynamics [22].

In the next section, we introduce definitions and notations useful for the subsequent discussion, following which in Section III we present an algorithm for inference of topology with unobserved nodes using inverse power spectral density. In Section IV, we present algorithms for exact topology learning of RLDS with partial observability, followed by results in Section V and conclusions in Section VI.

## II. PRELIMINARIES

Consider the continuous time linear dynamical system,

$$\sum_{m=1}^{l} a_{m,i}\frac{d^m x_i}{dt^m} = \sum_{j=1, j\neq i}^{n} b_{ij}(x_j(t) - x_i(t)) + p_i(t), \quad (1)$$

$i \in \{1, 2, .., n\}$, where, the exogenous forcing $p_i(t)$ is a zero mean wide sense stationary process uncorrelated with $p_j(t)$ for $j \neq i$. Here, $x_i(t) \in \mathbb{R}$ is a state of the system, $a_{m,i} \in \mathbb{R}$ and $b_{ij} \in \mathbb{R}_{\geq 0}$. Assuming that discrete time samples of the state $x_i$ are available as an output, we discretize the above

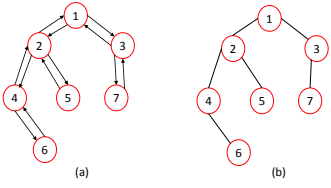

Fig. 1. (a) Graphical representation of a linear dynamical system where Assumption 1 holds, and (b) its associated topology.

continuous time dynamics using $z$ transform to obtain,

$$S_i(z)X_i(z) = \sum_{j=1, j\neq i}^{n} b_{ij}X_j(z) + P_i(z),$$

where, $S_i(z)$ is the frequency domain operator determined by the time derivatives of $x_i$. We rewrite the above equation as,

$$X_i(z) = \sum_{j=1, j\neq i}^{m} H_{ij}(z)X_j(z) + E_i(z) \quad (2)$$

where, $H_{ij}(z) = \frac{b_{ij}}{S_i(z)}, E_i(z) = \frac{1}{S_i(z)}P_i(z)$. Note that, for $j \neq i$, $e_i(k)$ are uncorrelated with $e_j(k)$ ($e_i$ is the inverse $z$ transform of $E_i(z)$ for all $i = 1,...,n$) and is a zero mean wide sense stationary sequence. Then, the dynamics of the entire network can be written as,

$X(z) = H(z)X(z) + E(z)$, where ,
$X(z) = [X_1(z) \ X_2(z) \ ... \ X_n(z)]^T$
$E(z) = [E_1(z) \ E_2(z) \ ... \ E_n(z)]^T, H(z)(i,j) = H_{ij}(z)$.

We assume that $I - H(z)$ is invertible almost everywhere. Since we are interested in bi-directed linear dynamical system, we make the following assumption in the rest of the paper.

**Assumption 1:** $H_{ij}(z) \neq 0$ almost surely implies $H_{ji}(z) \neq 0$ almost surely.

Assumption 1 is valid in linearized models of diverse engineering systems around an operating/ equilibrium point. For example - swing dynamics for power systems, lumped parameter RC network models for heat transfer dynamics and consensus networks. Note that the transfer functions $H_{ij}(z)$ and $H_{ji}(z)$ need not be same. We now associate a graphical model to the transfer function matrix $H(z)$.

**Graphical Representation:** Consider a directed graph $\mathcal{G} = (\mathcal{V}, \mathcal{E})$ with $\mathcal{V} = \{1,...,n\}$ and $\mathcal{E} = \{(i,j)|H_{ij}(z) \neq 0\}$. Each node $i \in \mathcal{V}$ is representative of the measured time series $x_i(k)$. In the graph $\mathcal{G}$, there is a directed edge from $j$ to $i$ if $H_{ij}(z) \neq 0$. It follows from Assumption 1 that, there is a directed edge from $i$ to $j$ as well. Thus $\mathcal{G}$ is a bi-directed graph. We call $\mathcal{G}$ to be the *generative graph* of the measured time series. Given generative graph $\mathcal{G}$, its *topology* is defined as the undirected graph $\mathcal{G}_T = (\mathcal{V}, \mathcal{E}_T)$ obtained by removing the orientation on all its edges, and avoiding repetition. An example of bi-directed generative graph and its topology are shown in Figure 1(a) and Figure 1(b) respectively. Next, we present terminology for undirected graphs which will be useful in the subsequent discussion.

*Definition 1:* (Path) A path between two nodes $x_0, x_k$ in an undirected graph $\mathcal{G}_T = (\mathcal{V}, \mathcal{E}_T)$ is a set of unique nodes $\{x_0, x_1, \cdots, x_k\} \subseteq \mathcal{V}$ where $\{(x_0, x_1), (x_1, x_2), \cdots, (x_{k-1}, x_k)\} \subseteq \mathcal{E}_T$. We will denote a path by $x_0 - x_1 - x_2 - \cdots - x_{k-1} - x_k$. The length of a path is one less than the number of nodes in the path. For example: $1 - 2 - 4 - 6$ is a path of length three between node 1 and 6 in the undirected graph of Figure 1(b).

*Definition 2:* ($n$ Hop Neighbor) In an undirected graph $\mathcal{G}_T = (\mathcal{V}, \mathcal{E}_T)$, $j \in \mathcal{V}$ is a $n$ hop neighbor of $i \in \mathcal{V}$, if there is a path of length $n$ between $i$ and $j$ in $\mathcal{G}_T$. For example: 1 and 6 are three hop neighbors in the undirected graph in Figure 1(b). If $n = 1$, $i$ and $j$ are termed neighbors in $\mathcal{G}_T$.

*Definition 3:* (Tree) A connected undirected graph without cycles is called a tree. There is a unique path between any two nodes in a tree.

*Definition 4:* (Leaf Node/ non-leaf Node of a Tree) In a tree $\mathcal{T}$, a node with degree 1 is called a leaf node. Nodes with degree greater than 1 are called non-leaf nodes.

Next we present the formal definition of a radial linear dynamical system (RLDS).

*Definition 5:* (Radial Linear Dynamical System) Consider a generative graph $\mathcal{G}$ with the associated topology being a tree, which is denoted by $\mathcal{T}$. A linear dynamical system with the above properties is referred to as a Radial Linear Dynamical System(RLDS). Figure 1(a) shows a RLDS with the corresponding topology $\mathcal{T}$ shown in Figure 1(b).

*Definition 6:* (Power Spectral Density(PSD) Matrix) For a $n$ dimensional collection of WSS time series $x(k) = \{x_1(k),...,x_n(k)\}^T$, the power spectral density matrix is defined as $\Phi_X(\omega) = \sum_{k=-\infty}^{\infty} E(x(k)x(0)^T)e^{-\iota \omega k}$.

In this article, we will focus on learning the topology of *radial linear dynamical systems* following Assumption 1. The only information available for topology estimation are time series measurements obtained from a subset of nodes in the system, while certain nodes are unobserved. Our analysis uses properties of inverse power spectral density of linear dynamical systems, which is presented next.

## III. TOPOLOGY LEARNING USING INVERSE PSD

Let $X(z) \in \mathbb{C}^n$ denote the vector of $z$ transform of $n$ nodal states, with $X(z) = [X_o(z)^T, X_h^T(z)]^T$, where, $X_o(z) \in \mathbb{C}^m$ and $X_h(z) \in \mathbb{C}^{n-m}$ are the $z$ transform of the nodal states corresponding to $m$ observed and $n-m$ unobserved nodes respectively. The network dynamics is represented in a compact form as,

$$\begin{bmatrix} X_o(z) \\ X_h(z) \end{bmatrix} = \begin{bmatrix} H_{oo}(z) & H_{oh}(z) \\ H_{ho}(z) & H_{hh}(z) \end{bmatrix} \begin{bmatrix} X_o(z) \\ X_h(z) \end{bmatrix} + \begin{bmatrix} E_o(z) \\ E_h(z) \end{bmatrix}$$

where, $E_o(z)$ and $E_h(z)$ denote the exogenous inputs at the observed and hidden nodes respectively. We assume that the unobserved nodes are not neighbors in $\mathcal{G}_T$, that is, $H_{hh}(z) = 0$. Let $\mathcal{V}_o$ denote the set of observed nodes and $\mathcal{V}_h$ denote the set of unobserved nodes and $\mathcal{V} = \mathcal{V}_o \cup \mathcal{V}_h$.

For notational simplicity we drop the argument $z$ in the discussion below. Let $\Phi_X$ denote the power spectral density matrix of the nodal states, that is,

$$\Phi_X = (I - H)^{-1}\Phi_E(I - H)^{-*}, \quad (3)$$

where, $\Phi_E$ is the diagonal matrix of power spectral densities of exogenous inputs and $*$ denotes the Hermittian operator. The objective of the following analysis is to show that inverse of the power spectral density of the states at the observed nodes (denoted by $\Phi_{oo}$) leads to a graph with spurious edges connecting up to four hop neighbors in $\mathcal{G}_T$. Let $J$ denote the inverse power spectral density matrix, that is,

$$J = \begin{bmatrix} J_{oo}(z) & J_{oh}(z) \\ J_{ho}(z) & J_{hh}(z) \end{bmatrix} = \Phi_X^{-1} = \begin{bmatrix} \Phi_{oo}(z) & \Phi_{oh}(z) \\ \Phi_{ho}(z) & \Phi_{hh}(z) \end{bmatrix}^{-1},$$
$$= (I - H(z))^* \Phi_E^{-1} (I - H(z)).$$

Using the matrix inversion lemma [23] it follows that,

$$\begin{align} \Phi_{oo}^{-1} &= J_{oo} - J_{oh} J_{hh}^{-1} J_{ho} \\ &=: \Gamma + \Delta + \Sigma \end{align} \quad (4)$$

where,

$$\begin{align} \Gamma &= (I - H_{oo}^*) \Phi_{E_o}^{-1} (I - H_{oo}), \\ \Delta &= H_{ho}^* \Phi_{E_h}^{-1} H_{ho}, \text{ and,} \\ \Sigma &= -\Psi^* \Lambda^{-1} \Psi, \text{ where} \\ \Lambda &= H_{oh}^* \Phi_{E_o}^{-1} H_{oh} + \Phi_{E_h}^{-1}, \\ \Psi &= H_{oh}^* \Phi_{E_o}^{-1} (I - H_{oo}) + \Phi_{E_h}^{-1} H_{ho}, \end{align} \quad (5)$$

*Lemma 3.1:* The following assertions hold

1) Suppose $i$ and $j$ are observed nodes and suppose in $\mathcal{G}_T$ (i) there is no path of the form $i - k - j$ with $k$ also observed and (ii) $i - j$ is not present, then $\Gamma(i, j) = 0$.
2) If in $\mathcal{G}_T$ there is no path between two observed nodes $i$ and $j$, connected via a single unobserved node, $k_u$, of the form $i - k_u - j$, then $\Delta(i, j) = 0$.
3) Suppose in $\mathcal{G}_T$ there is no path between two unobserved nodes with a single intermediate observed node; of the form $k_u - i - k'_u$ where $k_u$ and $k'_u$ are not observed and $i$ is observed, then $\Lambda$ is real and diagonal.
4) If in $\mathcal{G}_T$ for $j$ in the observed set of nodes and $k_u$ in the unobserved set of nodes; (i) $j - k_u$ is not present and (ii) there is no path of the form $j - p - k_u$ with $p$ being a observed node, then $\Psi(k, j) = 0$.
5) Suppose $\Lambda$ is diagonal, and if in $\mathcal{G}_T$, for observed nodes $i$ and $j$ and unobserved node $k_u$, there are no paths of the form $i - p - k_u$ or $i - k_u$ and $j - p' - k_u$ or $j - k_u$ for any $p$ and $p'$ being observed, then $\Sigma(i, j) = 0$.

*Proof:* 1) From (5),

$$\Gamma = \Phi_{E_o}^{-1} - \Phi_{E_o}^{-1} H_{oo} - H_{oo}^* \Phi_{E_o}^{-1} + H_{oo}^* \Phi_{E_o}^{-1} H_{oo}.$$

Note that $\Phi_{E_o}$ is diagonal for $i \neq j$; from which it follows that,

$$\begin{align} \Gamma(i, j) &= -\Phi_{E_o}^{-1}(i,i) H_{oo}(i,j) - \overline{H_{oo}(j,i)} \Phi_{E_o}^{-1}(j,j) \\ &+ \sum_{k=1}^{m} \overline{H_{oo}(k,i)} H_{oo}(k,j) \Phi_{E_o}^{-1}(k,k). \end{align}$$

The first two terms are zero if $i - j$ is not present in $\mathcal{G}_T$ and the third term is zero if a path of the form $i - k - j$ with $k$ being a observed node is not present in $\mathcal{G}_T$.

2) Note that $\Delta = H_{ho}^* \Phi_{E_h}^{-1} H_{ho}$ and thus for $i$ and $j$ in the observed set

$$\Delta(i, j) = \sum_{k_u \in \mathcal{V}_h} \overline{H_{ho}(k_u, i)} \Phi_{E_h}^{-1}(k_u, k_u) H_{ho}(k_u, j).$$

Thus if there is no path of the form $i - k_u - j$ where $k_u$ is unobserved, then $\Delta(i, j) = 0$.

3) Suppose $k_u \neq k'_u$ with $k_u$ and $k'_u$ being unobserved nodes. Note that $\Lambda(k_u, k'_u) = [H_{oh}^* \Phi_{E_o}^{-1} H_{oh} + \Phi_{E_h}^{-1}](k_u, k'_u)$. Thus $\Lambda(k_u, k'_u) = \sum_{i \in \mathcal{V}_o} H_{oh}^*(k_u, i) \Phi_{E_o}^{-1}(i,i) H_{oh}(i, k'_u) = \sum_{i \in \mathcal{V}_o} \overline{H_{oh}(i, k_u)} \Phi_{E_o}^{-1}(i,i) H_{oh}(i, k'_u)$, which is zero if there is no path of the from $k_u - i - k'_u$ with $i$ being an observed node. Moreover, $\Lambda(k_u, k_u) = \sum_{i \in \mathcal{V}_o} \overline{H_{oh}(i, k_u)} \Phi_{E_o}^{-1}(i,i) H_{oh}(i, k_u) + \Phi_{E_h}^{-1}(k_u, k_u) = \sum_{i \in \mathcal{V}_o} \Phi_{E_o}^{-1}(i,i) |H_{oh}(i, k_u)|^2 + \Phi_{E_h}^{-1}(k_u, k_u) \in \mathbb{R}$.

4) Note that

$$\begin{align} &\Psi(k_u, j) \\ &= [H_{oh}^* \Phi_{E_o}^{-1}](k_u, j) - [H_{oh}^* \Phi_{E_o}^{-1} H_{oo}](k_u, j) + [\Phi_{E_h}^{-1} H_{ho}](k_u, j) \\ &= \overline{H_{oh}(j, k_u)} \Phi_{E_o}^{-1}(j,j) - \sum_{p=1}^{m} \overline{H_{oh}(p, k_u)} \Phi_{E_o}^{-1}(p,p) H_{oo}(p, j) \\ &+ \Phi_{E_h}^{-1}(k_u, k_u) H_{ho}(k_u, j). \end{align} \quad (6)$$

The first and the last term are zero if $j - k_u$ is not present in $\mathcal{G}_T$ and the second term is zero if there exist no path of the form $j - p - k_u$ in $\mathcal{G}_T$, with $p$ being an observed node.

5) Note that if $\Lambda$ is diagonal, then,

$$\Sigma(i, j) = - \sum_{k_u \in \mathcal{V}_h} \overline{\Psi(k_u, i)} \Lambda^{-1}(k_u, k_u) \Psi(k_u, j),$$

Thus if there is no unobserved node $k_u$ with paths of the form $i - p - k_u$ or $i - k_u$ and $j - p' - k_u$ or $j - k_u$ for any $p$ and $p'$ being observed in $\mathcal{G}_T$, then from 4) of Lemma 3.1, $\overline{\Psi(k_u, i)} \Psi(k_u, j) = 0$ for every unobserved node $k_u$, which will imply $\Sigma(i, j) = 0$. This completes the proof. ∎

We use the above lemma to present a result on topology inference using the inverse of the power spectral density of the observed time series. In this regard we make the following assumption in the rest of the article.

**Assumption 2:** The unobserved nodes in topology $\mathcal{G}_T$ are at least four or more hops away from each other.

*Remark 1:* All the results presented in this article assume that the latent nodes are at least three or more hops away from each other except in Theorem 4.3, which requires that unobserved nodes are four or more hops away from each other.

*Theorem 3.1:* Consider a linear dynamical system with topology $\mathcal{G}_T$ such that Assumption 2 holds. Then $\Phi_{oo}^{-1}(i, j)(\omega) \neq 0$ almost surely for $\omega \in [0, 2\pi)$, implies that, $i$ and $j$ are within four hops of each other in the graph $\mathcal{G}_T$.

The proof follows from Lemma 3.1 and is omitted here due to space restriction.

*Remark 2:* Note that the non-zero values in $\Sigma(i, j)$ (and subsequently in $\Phi_{oo}^{-1}(i, j)$) for three and four hop observed nodes $i, j$ result from paths of the form $i - q - k_u - j, i - k_u - p - j$ and $i - q - k_u - p - j$ in $\mathcal{G}_T$, with $q$ and $p$ being observed neighbors of $i$ and $j$ respectively and $k_u$ being an unobserved node.

*Remark 3:* Note that the system transfer functions have to take very specific forms in order for $\Gamma(i, j) + \Delta(i, j) + \Sigma(i, j)$ to be zero even though $i, j$ are either neighbors or two hop neighbors. Thus, except for these pathological cases, if $i$ and $j$ are neighbors, two hop neighbors (with the common neighbor being observed or unobserved) $\Phi_{oo}^{-1}(i,j) \neq 0$ almost surely. Furthermore, for $i,j$ being three or four hop neighbors, $\Gamma(i,j) = 0$ and $\Delta(i,j) = 0$. The second term of $\Psi(i,j)$ is a contributor to three/ four hop contributions and is non zero in a large number of applications. For example: suppose that $b_{ij} \geq 0, a_{m,i} \geq 0$ for all $i,j$ in (1) (which is true for engineering networks like power distribution systems, RC networks etc.); then it is not possible that $\Phi_{oo}^{-1}(i,j) = 0$ if $i$ and $j$ are three or four hop neighbors in $\mathcal{G}_T$. We assume that the systems of interest do not belong to the small set of pathological cases.

If we form a graph $\mathcal{G}_m$ using the non-zero values in $\Phi_{oo}^{-1}(i,j)(\omega)$ as the adjacency matrix, we obtain all links up to four hop neighbors in $\mathcal{G}_T$. This is summarized as Algorithm 1 and is the first step in our topology learning scheme. The next objective is to identify the true links as well as eliminate the spurious links and identify the location of the unobserved nodes in $\mathcal{G}_m$ obtained from Algorithm 1. Note that Theorem 3.1 does not depend on the linear dynamical system being radial. However in the subsequent analysis, we will explicitly use the fact that $\mathcal{G}_T = \mathcal{T}$ is a RLDS.

---

**Algorithm 1** Topology Learning using Power Spectrum

---

**Input:** Time series $x_i(k)$ from observed nodes
**Output:** $\mathcal{G}_m = (\mathcal{V}_o, \mathcal{E}_{\mathcal{G}_m})$.
1: Edge set $\mathcal{E}_{\mathcal{G}_m} \leftarrow \{\}$
2: Compute $\Phi_{oo}^{-1}(\omega)$
3: **for all** $i \in \{1, 2, ..., m\}, i \neq j$ **do**
4:     **if** $\Phi_{oo}^{-1}(j,i)(\omega) \neq 0$ **then**
5:         $\mathcal{E}_{\mathcal{G}_m} \leftarrow \mathcal{E}_{\mathcal{G}_m} \cup \{(i,j)\}$
6:     **end if**
7: **end for**

---

## IV. EXACT TOPOLOGY RECOVERY IN RADIAL LINEAR DYNAMICAL SYSTEMS

To recover the exact topology of the radial linear dynamical system (RLDS) there the two tasks: one is to determine the set of true edges in the graph obtained from Algorithm 1 and the next is to determine the location of the unobserved nodes. We accomplish these tasks in the following two subsections.

### A. True Edge Discovery between Observed Nodes

Consider a RLDS with a tree topology $\mathcal{T}$. Let the unobserved nodes be at least four or more hops away from each other as per Assumption 2. The graph $\mathcal{T}_m$ obtained using Algorithm 1 has edges between observed nodes that are up to four hops away in $\mathcal{T}$. The objective of this section is to design an algorithm to identify the true links as well as eliminate the spurious links and detect the location of the missing nodes in $\mathcal{T}_m$ to recover $\mathcal{T}$ from $\mathcal{T}_m$. In this regard, we introduce the following assumption and the notion of separation in graphs.

**Assumption 3:** Each unobserved node is at least three hops away from all leaf nodes in $\mathcal{T}$.

Based on the above assumption, it is clear that all leaf nodes in $\mathcal{T}$ are observed nodes. Put differently, each unobserved node is buried deep into the network so that their effect is 'felt' at multiple observed nodes.

*Definition 7:* (Separation in Graph) In an undirected graph $\mathcal{U}$, the set of nodes $Z$ is said to separate the path between nodes $i$ and $j$, if there exist no path between $i$ and $j$ in $\mathcal{U}$ after removing the set of nodes $Z$. We will use the notation $sep(i, Z, j)$, which is to be read as $Z$ separates the path between $i$ and $j$ in $\mathcal{U}$. For example, in Figure 1(b) $sep(1, \{2, 4\}, 6)$ holds.

Next, we present a result, which enables us to categorize true and spurious edges between observed non-leaf nodes in $\mathcal{T}_m$. The proofs of the results presented below are omitted due to space restrictions.

*Theorem 4.1:* Consider a RLDS with a tree topology $\mathcal{T}$ such that Assumption 2 and 3 hold. Let $\mathcal{T}_m$ be such that there are links between any two observed nodes that are within four hops in the underlying topology $\mathcal{T}$ (that is no link up to four hops is undetected as discussed in Remark 3). There exist observed nodes $c, d$ distinct from observed nodes $a, b$ such that $sep(c, \{a, b\}, d)$ holds in $\mathcal{T}_m$ if and only if $a - b$ is an edge (and thus a true edge) in $\mathcal{T}$ and $a, b$ are non-leaf nodes.

*Remark 4:* The above theorem provides a topological test on $\mathcal{T}_m$ (which can be performed in polynomial time) to identify the observed non-leaf nodes, $\mathcal{V}_{nl,o}$ and the true edges between them. All other observed nodes in the graph $\mathcal{V}_l := \mathcal{V}_o \setminus \mathcal{V}_{nl,o}$ then are leaf nodes.

Note that of all edges connected to a leaf node in $\mathcal{T}_m$, only one edge is connected to its true non-leaf neighbor in $\mathcal{T}$ (rest are spurious edges). From Assumption 3, each leaf node is at least three hops away from any unobserved node in $\mathcal{T}$. By Lemma 3.1 and Remark 2, it clear that spurious edges connected with a leaf node in $\mathcal{T}_m$ include those to its two-hop neighbors. The next result utilizes the phase response of the entries of $\Phi_{oo}^{-1}$ to determine the true and spurious edges associated with leaf nodes.

*Theorem 4.2:* Consider a RLDS such that Assumption 2 and 3 hold. Let $a$ be a leaf node in $\mathcal{T}$ and let $v$ be a non-leaf neighbor of $a$ in $\mathcal{T}_m$. Then, $\angle \Phi_{a,v}^{-1}(\omega) = 0$ for all $\omega \in [0, 2\pi)$ if and only if $a, v$ are two hop neighbors in $\mathcal{T}$.

The proof uses algebraic expansions of the expressions for $\Phi_{a,v}^{-1}(\omega)$ for leaf $v$ and non-leaf $a$. We use the theorems mentioned above to devise Algorithm 2 that identifies all true edges between observed nodes in the system.

The last task that remains is to locate the unobserved nodes, which is discussed in the next subsection.

### B. Location of Unobserved Nodes

After application of Algorithm 1 followed by Algorithm 2, we end up with a graph $\overline{\mathcal{T}}$ of observed nodes and edges between them. However the discovered network will have multiple disconnected radial components, with the disconnections being at the locations of unobserved nodes. We will refer to $\overline{\mathcal{T}}_j$ as a **discovered disconnected component**. For example, consider a tree $\mathcal{T}$ with just one unobserved node $l$ as shown in Fig. 2. Let $l$ be between observed nodes $c, e$. Then there exist the path $C - c - l - e - E$ in $\mathcal{T}$, with

**Algorithm 2** True Edge Set Discovery Algorithm

**Input:** $\mathcal{T}_m = (V, \mathcal{E}_{\mathcal{T}_m})$ generated by Algorithm 1
**Output:** $\overline{\mathcal{T}} = (V, \mathcal{E}_{\overline{\mathcal{T}}})$
1: Edge set $\mathcal{E}_{\overline{\mathcal{T}}} \leftarrow \{\}$
2: **for all** edge $a - b$ in $\mathcal{E}_{\mathcal{T}_m}$ **do**
3:     **if** $Z := \{a, b\}$ there exist $I \neq \{\phi\}$ and $J \neq \{\phi\}$ such that $sep(I, Z, J)$ holds in $\mathcal{T}_m$ **then**
4:         $V_{nl} \leftarrow V_{nl} \cup \{a, b\}$, $\mathcal{E}_{\overline{\mathcal{T}}} \leftarrow \mathcal{E}_{\overline{\mathcal{T}}} \cup \{(a,b)\}$
5:     **end if**
6: **end for**
7: $V_l \leftarrow V - V_{nl}$
8: **for all** $a \in V_l, b \in V_{nl}$ with $(a,b) \in \mathcal{E}_{\mathcal{T}_m}$ **do**
9:     **if** $\angle \Phi_{oo}^{-1}(a,b) \neq 0$ for all $\omega \in [0, 2\pi)$ **then**
10:        $\mathcal{E}_{\overline{\mathcal{T}}} \leftarrow \mathcal{E}_{\overline{\mathcal{T}}} \cup \{(a,b)\}$
11:     **end if**
12: **end for**

$C := \{v \in \mathcal{V} | \text{path between } v, l \text{ involves node } c\}$,
$E := \{v \in \mathcal{V} | \text{path between } v, l \text{ involves node } e\}$.
Using Algorithm 2 leads to discovery of the individual components $C-c$ and $e-E$ in $\overline{\mathcal{T}}$. Based on our assumptions, it can be shown that each such component has at least three observed nodes. Since, $\mathcal{T}$ is a connected graph, the discovered disconnected components need to be connected by locating the unobserved node in $\overline{\mathcal{T}}$. Next, we present the result which enables us to do so.

*Theorem 4.3:* Let $\mathcal{T}_m$ be such that there are links between any two observed nodes that are within four hops in the topology $\mathcal{T}$. Consider two discovered disconnected components $\overline{\mathcal{T}}_1, \overline{\mathcal{T}}_2$ in $\overline{\mathcal{T}}$ with observed nodes $c \in \overline{\mathcal{T}}_1$ and $e \in \overline{\mathcal{T}}_2$. If $\forall\, b \in \overline{\mathcal{T}}_1, \forall f \in \overline{\mathcal{T}}_2$ such that $b-c$ and $e-f$ are edges in $\mathcal{T}$ and $b, c, e, f$ form a clique in $\mathcal{T}_m$, then, there exists an unobserved node $l$ such that $c-l-e$ is a path in $\mathcal{T}$.

Based on Theorem 4.3, we present Algorithm 3 that inserts hidden nodes by considering spurious edges between pairs of disconnected components in the discovered network. As each observed node can have a maximum of only one hidden node as neighbor, we merge hidden nodes that may have been duplicated in Algorithm 3 while checking for Theorem 4.3 between multiple disconnected components connected to the same hidden node.

**Algorithm 3** Unobserved Node Placement Algorithm

**Input:** $\overline{\mathcal{T}} = (\mathcal{V}_{\overline{\mathcal{T}}}, \mathcal{E}_{\overline{\mathcal{T}}}) = \cup_{j=1}^{h} \overline{\mathcal{T}}_j$
**Output:** $\tilde{\mathcal{T}} = (V_{\tilde{\mathcal{T}}}, \mathcal{E}_{\tilde{\mathcal{T}}})$.
1: Node set $V_{\tilde{\mathcal{T}}} \leftarrow \mathcal{V}_{\overline{\mathcal{T}}}$
2: Edge set $\mathcal{E}_{\tilde{\mathcal{T}}} \leftarrow \mathcal{E}_{\overline{\mathcal{T}}}$
3: **for all** $j \in \{1, 2, ..., h\}$ **do**
4:     **for all** $i \in \{j+1, ..., h\}$ **do**
5:         **if** there exist a pair of nodes $a, b$ such that $a \in \overline{\mathcal{T}}_j$ and $b \in \overline{\mathcal{T}}_i$ such that all their neighbors in $\overline{\mathcal{T}}$ are connected in $\mathcal{T}_m$ **then**
6:             $V_{\tilde{\mathcal{T}}} \leftarrow V_{\tilde{\mathcal{T}}} \cup l_j$
7:             $\mathcal{E}_{\tilde{\mathcal{T}}} \leftarrow \mathcal{E}_{\tilde{\mathcal{T}}} \cup \{(a, l_j), (l_j, b)\}$
8:         **end if**
9:     **end for**
10: **end for**
11: Merge hidden nodes that are neighbors of the same observed node.

## V. RESULTS

Topology inference for power distribution networks can be applied towards fault isolation, control and flow opti-mization. Penetration of devices like Phasor Measurement Units(PMUs) enable real time measurement of phases of various nodes and facilitate inverse problems like topology inference and state estimation [24]. However, these meters cannot be deployed at all nodes and partial network observability is indeed the situation.

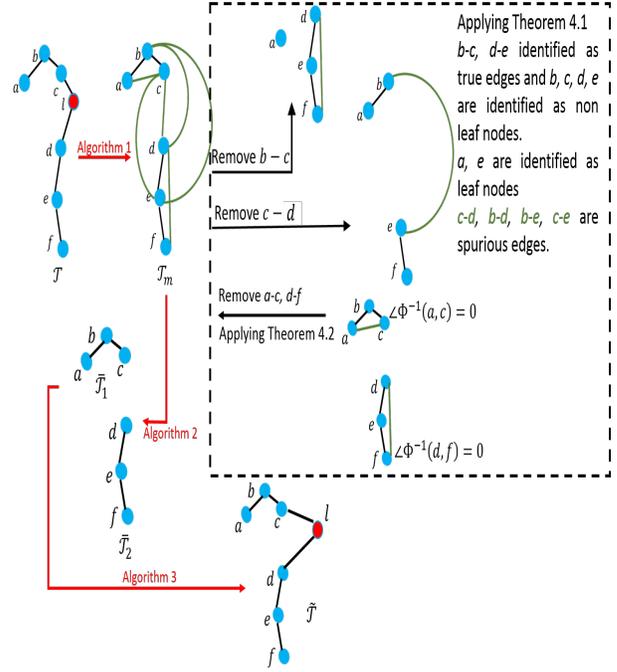

Fig. 2. Illustration of the application of Algorithm 1, Algorithm 2 and Algorithm 3 in succession. The red node is the latent node and green edges denote the spurious edges up to four hop neighbors.

We demonstrate the efficacy of the algorithm presented by testing it on data obtained from simulations of the linearized swing equations (see (7) below) on a 39 bus radial topology. This radial system is obtained by deleting a few edges from the IEEE 39 bus system and is shown in Figure 3(a). The state $x_i(t)$ denotes the fluctuations of the phase angles of node $i$ from equilibrium values while $p_i(t)$ denote the nodal injections due to generation and losses. The nodal injections $p_j$ are colored noise and are generated by filtered version of a white noise sequence. Four nodes are unobserved in this study. For $i = 1, 2, ..., 39$,

$$m_i \ddot{x}_i + d_i \dot{x}_i = \sum_{j=1, j \neq i}^{39} b_{ij}(x_j(t) - x_i(t)) + p_i(t), \quad (7)$$

Here, $m_i, d_i, b_{ij} \in \mathbb{R}_{\geq 0}$ for all $i, j \in \{1, 2, ..., 39\}$.

The error proportion is defined as the ratio of the sum of number of true links undetected and number of false links detected to the total number of true links. The error proportion for the RLDS in Figure 3(a) using the algorithms presented previously is shown in Figure 3(b). As the samples per observed node is increased it is seen that the error proportion decreases rapidly. We reemphasize that our algorithms do not use any knowledge of the system parameters and noise injections. Moreover, the noise injections used in the simulation is colored noise unlike white noise models used in previous studies.

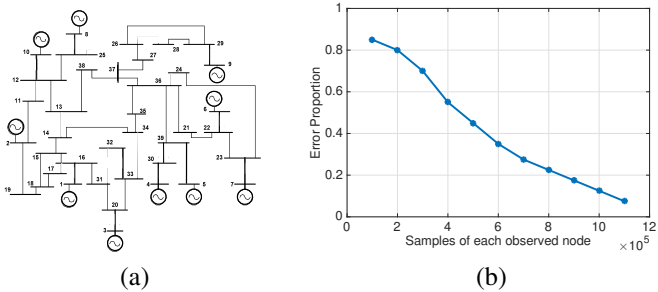

(a)          (b)

Fig. 3. (a) A RLDS obtained from the IEEE 39 bus system, (b) error proportion against the number of samples in topology inference of the system shown in Figure 2(a).

## VI. Conclusion

We presented algorithms, which when applied in succession, leads to the exact topology recovery of a RLDS in the 'sufficient statistics'(large number of data samples) regime under partial observation of the network. The proofs involve a synergy of tools from signal processing and probabilistic graphical models. Algorithm 2 and Algorithm 3 being graph based checks, can be executed in polynomial time. Among all the algorithms presented, Algorithm 1 is computationally most intensive due to computation of the inverse. We demonstrated the performance of the algorithm on a 39 node radial power distribution network. This work also provides insights on placement of sensors for observing the network for monitoring and fault detection applications.

In future, we plan to relax the assumptions of unobserved nodes being at least four hops away from each other as well as at least three hops away from any leaf node.